\documentclass[journal=jacsat,manuscript=article]{achemso}
\usepackage[version=3]{mhchem} 
\usepackage{xcolor}
\usepackage{graphicx}
\usepackage{subcaption}
\usepackage{soul}
\usepackage{float} 

\author{Y. Ricardo Espinosa}
\affiliation{Institute of Physics of Liquids and Biological Systems (IFLYSIB), CONICET, La Plata, Argentina.}
\alsoaffiliation{Grupo de Investigaci\'on en Biolog\'ia Molecular y Gen\'etica, Facultad de Ciencias B\'asicas, Universidad de Pamplona, Pamplona, Colombia.}

\author{C. Manuel Carlevaro}
\affiliation{Institute of Physics of Liquids and Biological Systems (IFLYSIB), CONICET, La Plata, Argentina.}
\alsoaffiliation{Centro de Investigaci\'on en Mec\'anica Experimental y Computacional (CIMEC), Universidad Tecnol\'ogica Nacional, Facultad Regional La Plata, Berisso, Argentina.}

\author{C. Gast\'on Ferrara}
\email{gastonf@iflysib.unlp.edu.ar}
\phone{+54 (0)221 4254904}
\affiliation{Institute of Physics of Liquids and Biological Systems (IFLYSIB), CONICET, La Plata, Argentina.}
\alsoaffiliation{Institute of Engineering and Agronomy, National University Arturo Jauretche, Av.\ Calchaqu\'i 6200, Florencio Varela, Argentina.}

\title[Urea-Mediated Solvent Reorganization of BSA at pH~3.7]
  {Urea-Mediated Solvent Reorganization of Bovine Serum Albumin in an Acid-Induced Expanded Conformation at pH~3.7}

\abbreviations{BSA,MD,RDF,HB,SASA,RG,RMSD,COM}

\begin{document}

\begin{abstract}
	Understanding the molecular mechanisms by which denaturants modulate protein structure remains a central challenge in protein biophysics. In this study, molecular dynamics (MD) simulations were employed to investigate the effects of urea on the structural stability of bovine serum albumin (BSA) in its F isoform at pH~3.7 across a broad range of urea concentrations, from pure aqueous solution (0~M) to a fully urea-solvated environment. The simulations reveal a concentration-dependent remodeling of the protein hydration shell. At low urea concentrations, backbone--water hydrogen bonds decrease by approximately 40\%, accompanied by an approximately 45\% increase in protein--urea hydrogen bonds between 1 and 7~M urea, consistent with a competitive solvation process in which urea progressively replaces water molecules at the protein surface. As urea concentration increases, urea--urea self-association becomes increasingly significant, reducing the number of direct protein--urea contacts; concurrently, the remaining water molecules form protein--water hydrogen bonds more efficiently on a per-water-molecule basis, without implying a net increase in the absolute number of hydration water molecules.

	Despite these pronounced solvent rearrangements, the secondary structure of BSA remains largely preserved throughout the simulations. In contrast, local structural organization and global conformational features, particularly within Domain~III, exhibit increased solvent exposure and enhanced conformational flexibility. These observations support a dynamic compensation mechanism in which urea partially substitutes water molecules within the hydration shell without fully disrupting the underlying hydrogen-bond network that stabilizes the protein. Overall, this study provides molecular-level insight into the interplay between preferential interactions, solvation dynamics, and protein stability under denaturing conditions.

\end{abstract}
\section{Introduction}

Proteins function within a highly dynamic aqueous environment in which interactions with the surrounding solvent critically govern structural stability, conformational flexibility, and biological activity. Water actively contributes to the stabilization of secondary and tertiary structures, modulates conformational fluctuations, and plays a central role in the thermodynamic balance that defines the ensemble of accessible conformational states \cite{kazlauskas2018engineering,gazi2023conformational,lim2009urea,biswas2018contrasting}. Consequently, variations in solvent composition can substantially alter this equilibrium by modifying hydration patterns and protein--solvent interactions, thus favoring or destabilizing specific conformational states.

Urea is perhaps the most extensively studied protein denaturant; nevertheless, despite decades of experimental and theoretical investigations, the molecular basis of its denaturing action remains incompletely understood \cite{khan2019protein,paladino2023action,canchi2010equilibrium,almarza2009molecular}. Traditionally, two main mechanisms have been proposed to explain urea-induced protein unfolding: an indirect mechanism, whereby urea perturbs the structure and thermodynamic properties of water, thereby weakening hydrophobic interactions, and a direct mechanism involving favorable interactions between urea molecules and the protein surface \cite{khan2019protein}. Although current experimental and computational studies indicate that both mechanisms may contribute to the unfolding process, accumulating evidence supports a predominant role of direct protein--urea interactions in stabilizing unfolded conformational ensembles \cite{paladino2022structure,canchi2010equilibrium,paladino2023action}.

The denaturing activity of urea is generally attributed to its remarkable ability to interact with a wide range of protein chemical groups. Owing to its high polarizability, its capacity to both donate and accept hydrogen bonds, and its favorable dispersion interactions with peptide backbones and amino acid side chains, urea can efficiently associate with both polar and hydrophobic regions of proteins \cite{paladino2023action,nakata2023molecular}. Molecular dynamics simulations and thermodynamic analyses have consistently shown that these preferential interactions favor conformations with larger solvent-accessible surface areas, thereby shifting the equilibrium toward more expanded and unfolded states \cite{canchi2010equilibrium,nakata2023molecular}. Nevertheless, urea-induced denaturation cannot be interpreted solely in terms of direct protein--urea contacts. Growing evidence indicates that unfolding is accompanied by substantial changes in the organization and dynamics of the hydration shell, in which competition between water and urea molecules for the protein surface leads to a complex reorganization of local solvation environments. As a result, both the conformational landscape of the protein and the physicochemical properties of the surrounding solvent are modified in a highly coupled manner \cite{qausain2020mechanistic,ganguly2019distinct,gazi2023conformational}.

Increasing urea concentration profoundly alters the composition and organization of the protein hydration shell. As urea accumulates near the protein surface, water molecules are progressively redistributed, leading to a reduction in protein--water hydrogen bonding and a corresponding increase in direct protein--urea interactions \cite{qausain2020mechanistic}. Nevertheless, the resulting changes cannot be described simply as a net loss of hydration water. Recent studies have demonstrated that the response of the hydration environment depends strongly on the local chemical character of the protein surface, the conformational state of the protein, and the cosolvent concentration \cite{ganguly2019distinct,hishida2023contrasting, guckeisen2021effect}. Under these conditions, water depletion, restructuring of hydrogen-bond networks, and localized rehydration phenomena may occur concurrently, highlighting that urea-induced conformational changes emerge from a collective reorganization of the protein--water--cosolvent environment rather than from a simple replacement of water molecules at the protein surface.

The importance of these processes becomes particularly evident in partially unfolded states. While much of our current understanding of protein denaturation has been derived from comparisons between native and fully unfolded conformations, many biological proteins populate intermediate states in which structurally preserved regions coexist with more flexible domains that are increasingly exposed to the solvent \cite{paladino2023action,gazi2023conformational}. Such conformational ensembles provide an ideal framework for investigating how competition between water and cosolvents modulates protein stability before the onset of global structural disruption.

Bovine serum albumin (BSA) represents a particularly suitable model system for investigating these phenomena owing to its abundance, structural stability, and extensive experimental and computational characterization \cite{munoz2022characterization,jia2025insights,nnyigide2018exploring}.

Under acidic conditions, BSA transitions to the so-called F form, an expanded conformational state in which a substantial fraction of the secondary structure is retained. At the same time, significant alterations occur in tertiary organization and interdomain interactions. The increased exposure of regions that are normally buried within the protein interior renders this conformational state especially sensitive to perturbations in the hydration environment, thereby providing a valuable framework for elucidating the molecular mechanisms underlying urea-induced solvent reorganization.

In this work, we employ all-atom molecular dynamics simulations to investigate the effect of urea on solvent organization and structural stability in bovine serum albumin (BSA) adopting its acid-induced expanded conformation at pH 3.7. By characterizing the interplay between protein structure, hydration, and protein--urea interactions, this study provides molecular-level insights into the mechanisms governing urea-induced perturbations in partially unfolded proteins. The agreement between our findings and previously reported experimental and computational observations further supports the reliability of our approach and advances the current understanding of urea-mediated protein denaturation.

\section{Materials and Methods}

\subsection{Systems and Simulations}

The effects of urea on the structural stability of bovine serum albumin (BSA) were investigated in aqueous solutions spanning a broad range of urea concentrations. Urea molecules were modeled using the Boek parameterization \cite{boek1994interfaces}, while bonded and nonbonded interaction parameters were described using the GROMOS 54A7 force field \cite{schmid2011definition}.

Water molecules were represented using the extended simple point charge (SPC/E) model \cite{berendsen1987missing}.
The initial atomic coordinates of bovine serum albumin (BSA) were obtained from its crystallographic structure deposited in the Protein Data Bank (PDB ID: 4F5S) \cite{bujacz2012structures}. The protonation states and corresponding protein topology representative of acidic conditions at pH~3.7 were adopted from previous studies combining Small-Angle X-ray Scattering (SAXS) experiments and molecular dynamics simulations \cite{scanavachi2020aggregation}. Under these conditions, BSA carries a net charge of $+99\,e$, which is neutralized by 99 chloride ions to ensure the overall electroneutrality of the system.
All intermolecular interactions were treated explicitly using the \texttt{GROMACS} 2023.4 molecular dynamics simulation package \cite{spoelgromacs}. As a reference condition, a control system in the absence of urea was initially constructed, in which water molecules were randomly distributed around the protein (\emph{Control W}, Table~\ref{table1}).


\begin{table}[H]
	\centering
	\caption{Composition of the simulated systems used to investigate the effects of urea on the structural stability of bovine serum albumin (BSA). The systems include a pure aqueous control (\emph{Control W}), urea--water mixtures at concentrations ranging from 1 to 7~M, and a fully urea-solvated system (\emph{Control U}).}
	\label{table1}

	\begin{tabular}{lcccc}
		\hline
		\textbf{System}          &
		\textbf{Water Molecules} &
		\textbf{Counterions}     &
		\textbf{Urea Molecules}  &
		\textbf{Urea Concentration}                                 \\
		\hline
		\emph{Control W}         & 61,240 & 99 & 0      & 0~M       \\
		1~M                      & 57,321 & 99 & 1,099  & 1~M       \\
		2~M                      & 53,126 & 99 & 2,198  & 2~M       \\
		3~M                      & 49,168 & 99 & 3,278  & 3~M       \\
		4~M                      & 44,997 & 99 & 4,396  & 4~M       \\
		5~M                      & 40,964 & 99 & 5,468  & 5~M       \\
		7~M                      & 32,412 & 99 & 7,570  & 7~M       \\
		\emph{Control U}         & 0      & 99 & 19,774 & neat urea \\
		\hline
	\end{tabular}
\end{table}

The control system was equilibrated using a two-step protocol. Initially, a 500~ps simulation in the NVT ensemble was performed with harmonic positional restraints applied to the protein C$\alpha$ atoms, using a force constant of 1000~kJ~mol$^{-1}$~nm$^{-2}$ along each Cartesian direction. The positional restraints were subsequently removed, and the system was further equilibrated for 500 ps in the NpT ensemble to allow for density and pressure stabilization.

Following equilibration of the pure aqueous control system (\emph{Control W}), the resulting configuration was used as the initial structure for all urea-containing simulations. Urea molecules were then randomly inserted into the solvent phase, generating six urea--water systems spanning concentrations from 1 to 7~M, together with a fully urea-solvated system (\emph{Control U}) (Table~\ref{table1}). All systems listed in Table~\ref{table1} were equilibrated using the same two-step protocol described for the control simulation, ensuring methodological consistency across all conditions.

\begin{figure*}[ht!]
	\centering
	\includegraphics{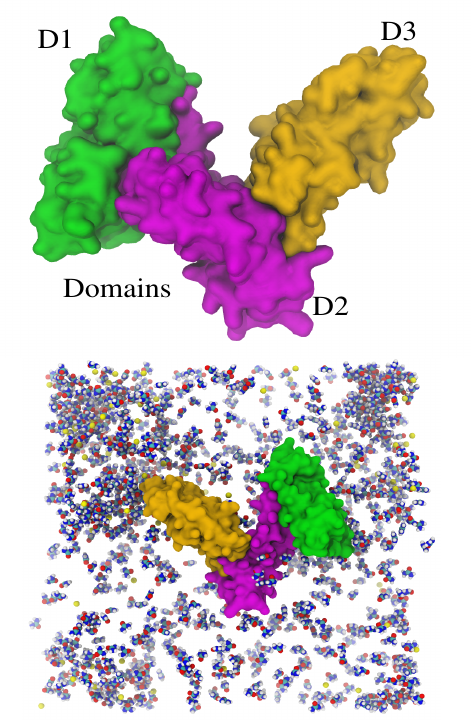}
	\caption{Molecular representation of bovine serum albumin (BSA) in a 1~M urea-containing solvent box. BSA is shown divided into three structural domains: Domain~I (D1, green; residues 1--185), Domain~II (D2, magenta/violet; residues 186--378), and Domain~III (D3, yellow/gold; residues 379--576). The upper panel highlights the domain organization of BSA, whereas the lower panel shows the protein embedded in the simulation box containing urea molecules, water molecules, and chloride ions. Water molecules are not shown for clarity. Chloride ions are represented as yellow spheres, while urea molecules are shown using cyan carbon, red oxygen, blue nitrogen, and white hydrogen atoms.}
	\label{Fig1}
\end{figure*}

Subsequently, 500~ns production simulations were carried out for each system. Figure~\ref{Fig1} illustrates the 1~M system following 500 ns of production MD simulation.

To improve conformational sampling, ten configurations were randomly extracted from the final 10~ns of each production trajectory and used as starting points for independent 10~ns simulations, each initialized with a distinct velocity distribution. This approach enabled a more extensive exploration of long-timescale conformational dynamics and intermolecular interactions within the protein--urea--water systems.

All equilibration and production simulations were performed in cubic simulation boxes under periodic boundary conditions at 300.15~K and 1~bar. During the equilibration stages, temperature and pressure were controlled using the velocity-rescaling thermostat (V-rescale) \cite{bussi2007canonical} and the Berendsen barostat \cite{berendsen1984molecular}, with coupling constants of 0.1~ps and 1.0~ps, respectively. Production simulations were subsequently carried out using the velocity-rescaling thermostat (V-rescale) together with the canonical sampling through velocity rescaling barostat (C-rescale), ensuring improved ensemble fluctuations during the production stage. Long-range electrostatic interactions were treated using the particle mesh Ewald (PME) method \cite{darden1993particle,essmann1995smooth,abraham2011optimization}. A cutoff distance of 1.0~nm was applied to both van der Waals and Coulomb interactions. Covalent bonds involving solute atoms were constrained using the LINCS algorithm \cite{hess1997lincs}, allowing the use of a 2~fs integration time step in all simulations.

\textbf{Hydrogen Bonds.} Hydrogen-bond populations were calculated using the \texttt{gmx hbond} tool implemented in GROMACS. Hydrogen bonds were identified according to geometric criteria, considering a donor--acceptor distance of $\leq 0.35$~nm and an acceptor--donor--hydrogen (A--D--H) angle of $\leq 30^\circ$, consistent with the standard hydrogen-bond definition employed in GROMACS.

\section{Results and Discussion}

All analyses presented in this section are based on ensemble averages obtained from ten independent simulations of 10~ns each, generated for every system investigated, including the pure aqueous control (\emph{Control W}), urea--water mixtures spanning concentrations from 1 to 5~M and 7~M, and the fully urea-solvated system (\emph{Control U}). Before data acquisition, all systems were subjected to 500~ns production simulations to ensure adequate hydration, solvation, and structural equilibration of the protein within each solvent environment. This sampling strategy was designed to improve the statistical robustness of the analyses and to provide a more comprehensive characterization of the intermolecular interactions and conformational behavior of BSA across the different denaturing conditions.

\subsection{Solvent-Mediated Hydrogen-Bond Remodeling in BSA}

\begin{figure*}[ht!]
	\centering
	\includegraphics[width=0.7\linewidth]{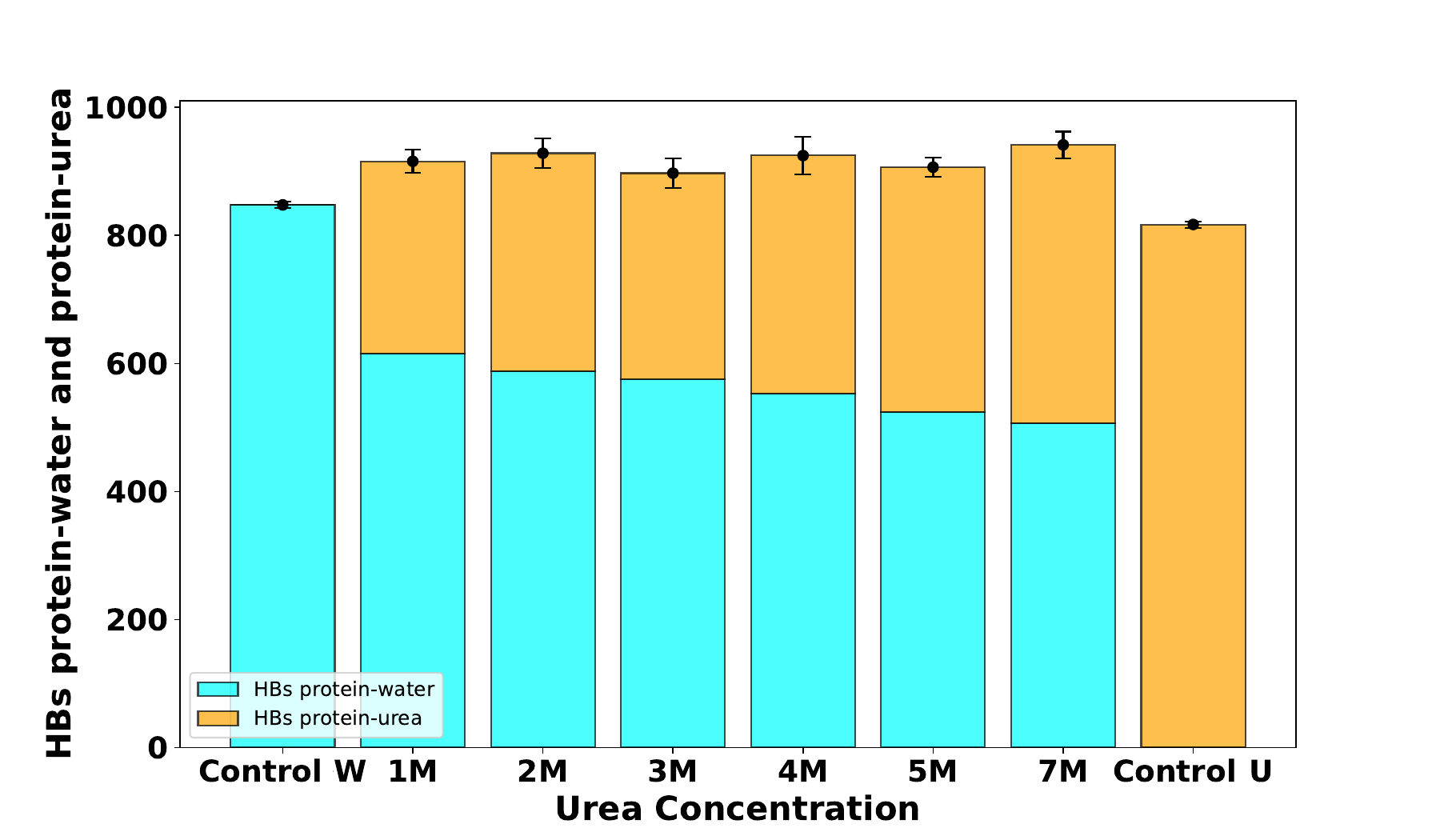}
	\caption{Average number of hydrogen bonds (HBs) formed between BSA and solvent components as a function of urea concentration. Protein--water and protein--urea HBs are shown separately, together with their combined contribution.}
	\label{Fig2}
\end{figure*}

Hydrogen-bond formation between BSA and the surrounding solvent was analyzed as a function of urea concentration. For each condition, hydrogen-bond populations were averaged over the final 8 ns of ten independent simulations. Figure~\ref{Fig2} shows the average number of protein--water and protein--urea hydrogen bonds (HBs), together with their combined contribution, and compares these values with the urea-free reference system.

A progressive reduction in protein--water hydrogen bonds was observed as the urea concentration increased, indicating a gradual disruption of the native hydration shell surrounding BSA. This decrease was accompanied by a concomitant increase in protein--urea hydrogen bonds, which increased by approximately 45\% between 1 and 7~M urea. These results suggest that urea progressively accumulates within the protein solvation shell and competes with water for hydrogen-bonding sites on the protein surface.

Interestingly, the combined number of protein--water and protein--urea hydrogen bonds in the urea--water mixtures exceeded the number of protein--water hydrogen bonds observed in the urea-free control. However, this increase in total solvent-mediated hydrogen bonding should not be interpreted as a direct indication of enhanced structural stability. Rather, it reflects a remodeling of the protein solvation environment, in which water-mediated interactions are progressively replaced or supplemented by urea-mediated contacts.

The comparison between the pure water and pure urea control systems further indicates that urea does not reproduce the hydrogen-bonding pattern established by water around BSA. Although urea can partially compensate for the loss of protein--water hydrogen bonds, it generates a chemically distinct solvation network. This altered interaction pattern may perturb the native hydration shell, modify solvent exposure of polar and backbone groups, and contribute to changes in protein stability and structural integrity under denaturing conditions.

\begin{figure}[htbp]
	\centering
	\includegraphics[width=0.99\textwidth]{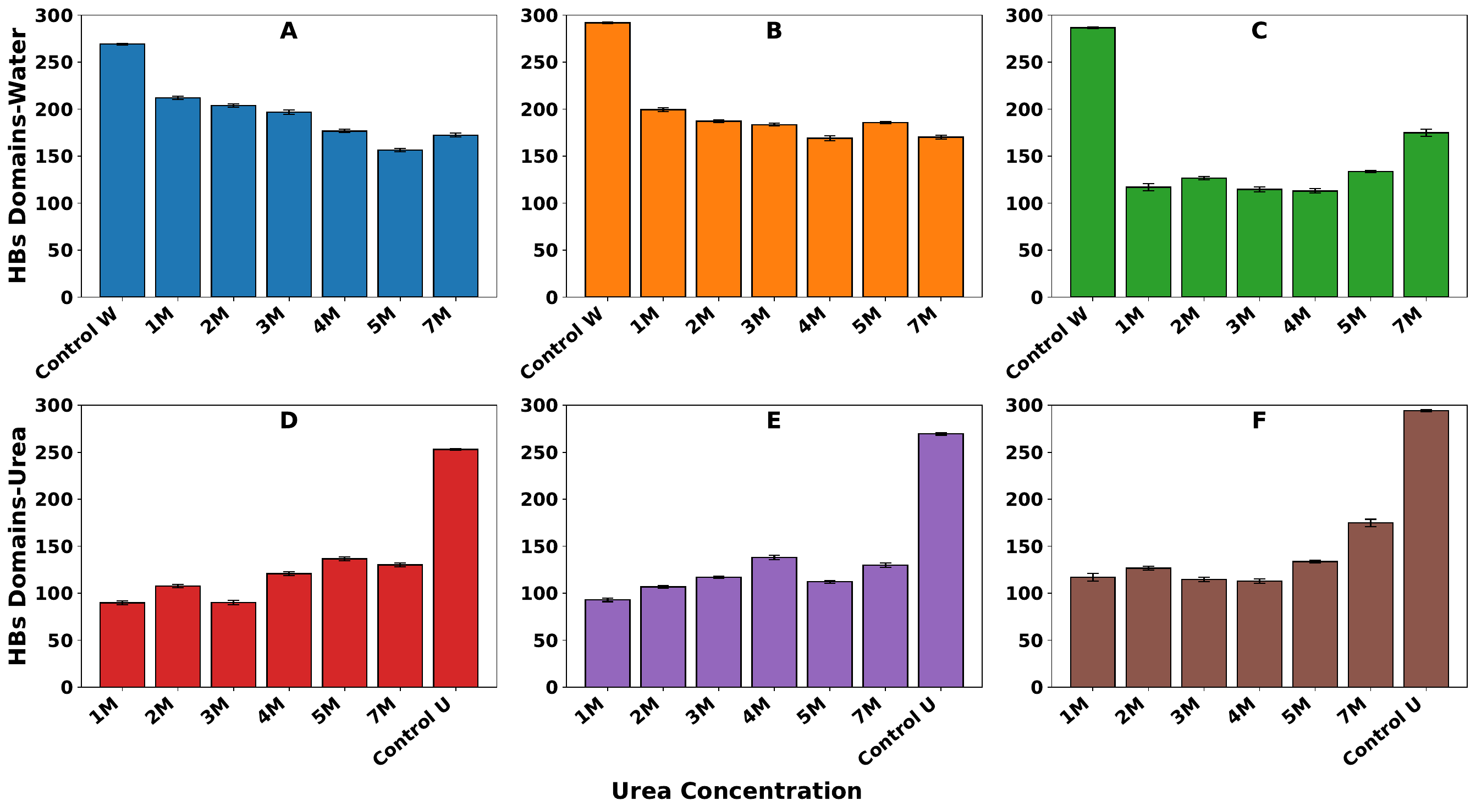}
	\caption{Domain hydrogen bond analysis between BSA and solvent components as a function of urea concentration. The upper row shows the number of hydrogen bonds formed between water and Domains~I (A), II(B), and III(C), respectively. The lower row shows the number of hydrogen bonds formed between urea and Domains~I(D), II(E), and III(F), respectively.}
	\label{Fig3}
\end{figure}

At the domain level, the analysis of protein--water hydrogen bonds revealed a progressive reduction in hydration-mediated interactions as the urea concentration increased (Figure~\ref{Fig3}, upper row). This trend was observed in all three domains, with the most pronounced decreases occurring at high urea concentrations, particularly at 5 and 7~M.

Although all domains displayed comparable protein--water hydrogen-bond populations under the 0 and 1~M conditions, differences became more evident as the urea concentration increased. Domain~III showed a relatively conserved hydrogen-bonding pattern at low urea concentrations, suggesting that this domain may initially retain its hydration environment more effectively than Domains~I and II. However, at higher urea concentrations, this protective hydration pattern was progressively lost, indicating that the effect of urea becomes more generalized across the protein structure under strongly denaturing conditions.

The analysis of domain--urea hydrogen bonds showed the opposite behavior (Figure~\ref{Fig3}, lower row). As the urea concentration increased, all three domains exhibited a progressive increase in the number of hydrogen bonds formed with urea molecules.

Taken together, these results indicate that urea-induced disruption of the BSA hydration shell is a domain-dependent process. The progressive loss of protein--water hydrogen bonds, together with the concomitant gain in protein--urea hydrogen bonds, suggests that urea replaces water in the local solvation environment of the protein. However, the magnitude of this replacement varies across domains, suggesting that local solvent accessibility, surface polarity, and domain compactness modulate each region's susceptibility to urea-mediated perturbation.

Backbone- and side-chain-resolved analyses revealed that urea affects these regions differently. The addition of urea produced a marked reduction in backbone--water hydrogen bonds, reaching approximately 40\% relative to the urea-free system, while backbone--urea hydrogen bonds increased with urea concentration and became particularly abundant in the pure urea control. This behavior suggests that urea directly competes with water for hydrogen-bonding sites along the protein backbone. In contrast, side-chain hydrogen bonding with the solvent was comparatively more conserved, as the reduction in side-chain--water interactions was partially compensated by the formation of side-chain--urea hydrogen bonds. Thus, the backbone appears to be more sensitive to urea-mediated dehydration, whereas solvent-exposed polar and charged side chains retain a relatively stable hydrogen-bonding capacity through replacement of water by urea (see Figures~S1 and~S2 in the Supporting Information).

\subsection{Global and Domain-Resolved Structural Response of BSA to Urea}

All structural descriptors presented in this section were computed from the final 8 ns of the ten replica simulations performed for each urea concentration. The reported values represent time-averaged quantities subsequently averaged across replicas for each condition.

\begin{figure}[htbp]
	\centering
	\includegraphics[width=0.65\textwidth]{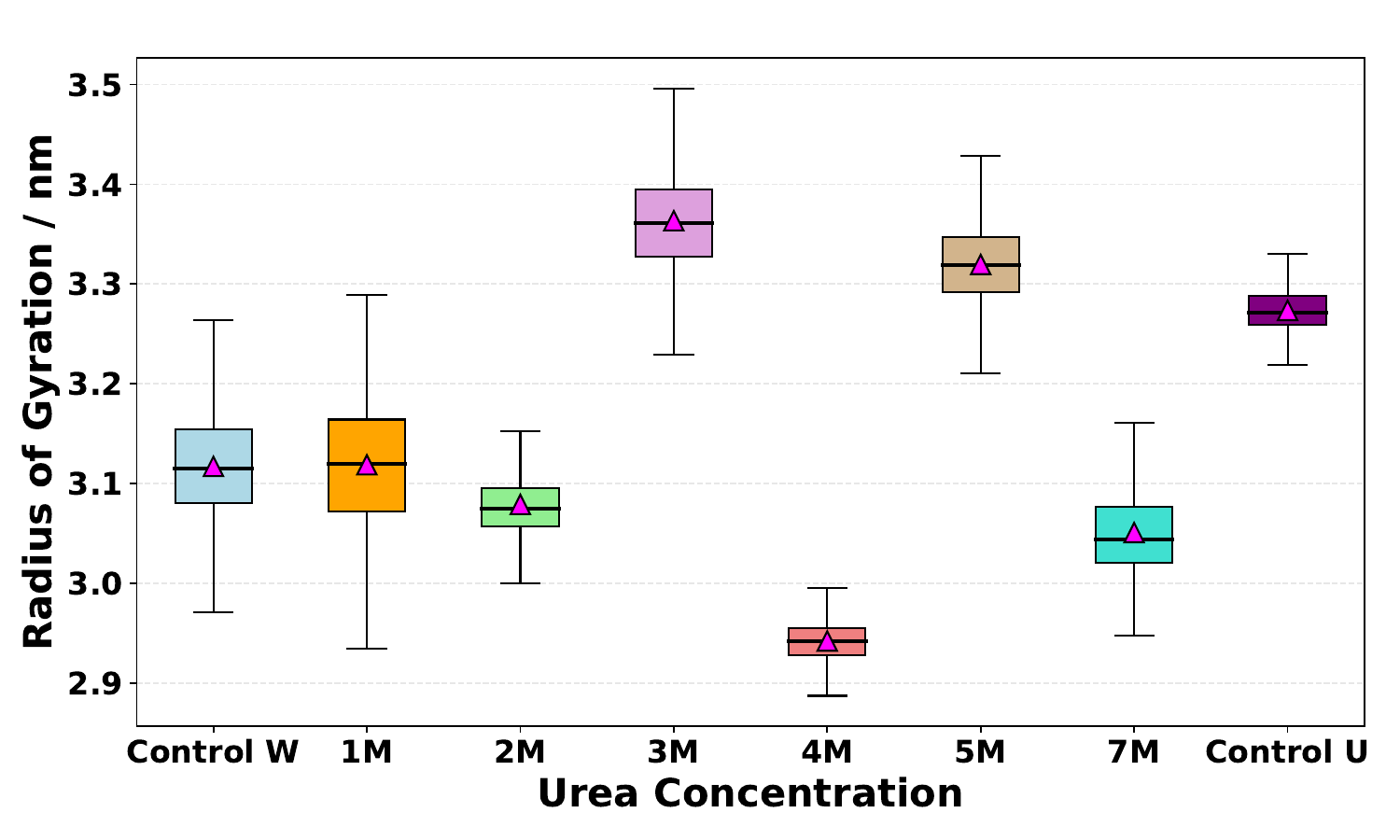}
	\caption{Radius of Gyration (RG) vs urea concentration.}
	\label{Fig4}
\end{figure}

In Figure~\ref{Fig4}, the radius of gyration of BSA is analyzed as a function of urea concentration. At low urea concentrations, the average radius of gyration remains nearly constant, indicating that urea does not strongly perturb the hydration shell surrounding the protein backbone. However, at higher urea concentrations, the average radius of gyration exhibits moderate fluctuations without following a strictly monotonic trend. This behavior suggests that urea induces only limited changes in the global compactness of BSA rather than promoting a progressive concentration-dependent expansion.

In this case, the presence and increase of urea does not consistently increase the average radius of gyration relative to the urea-free system \cite{espinosa2025molecular}. Moreover, the observed variations are modest, not exceeding 8\%, suggesting that any structural changes remain limited in magnitude. At low urea concentrations, these findings are in agreement with the results reported by Hayashi and co-workers \cite{hayashi2007protein}, who, within the 0--5 M range and at the same temperature, observed that the protein structure remains largely preserved, with only minor fluctuations associated with solvent dynamics.

At high urea concentrations (7 M and Control U), no substantial variations in the radius of gyration are detected. Unlike the behavior reported by Hasan and co-workers \cite{ref_hasan}, our simulations do not reveal pronounced side-chain expansion in BSA nor a marked increase in hydrophobic hydration effects. This difference may be largely attributed to the initial structural model adopted in our study, which reflects acidic conditions at pH 3.7. Under these conditions, the protein is already partially unfolded, potentially limiting the extent of additional conformational changes induced by urea \cite{atahar2019aggregation}.

\begin{figure}[htbp]
	\centering
	\includegraphics[width=0.47\textwidth]{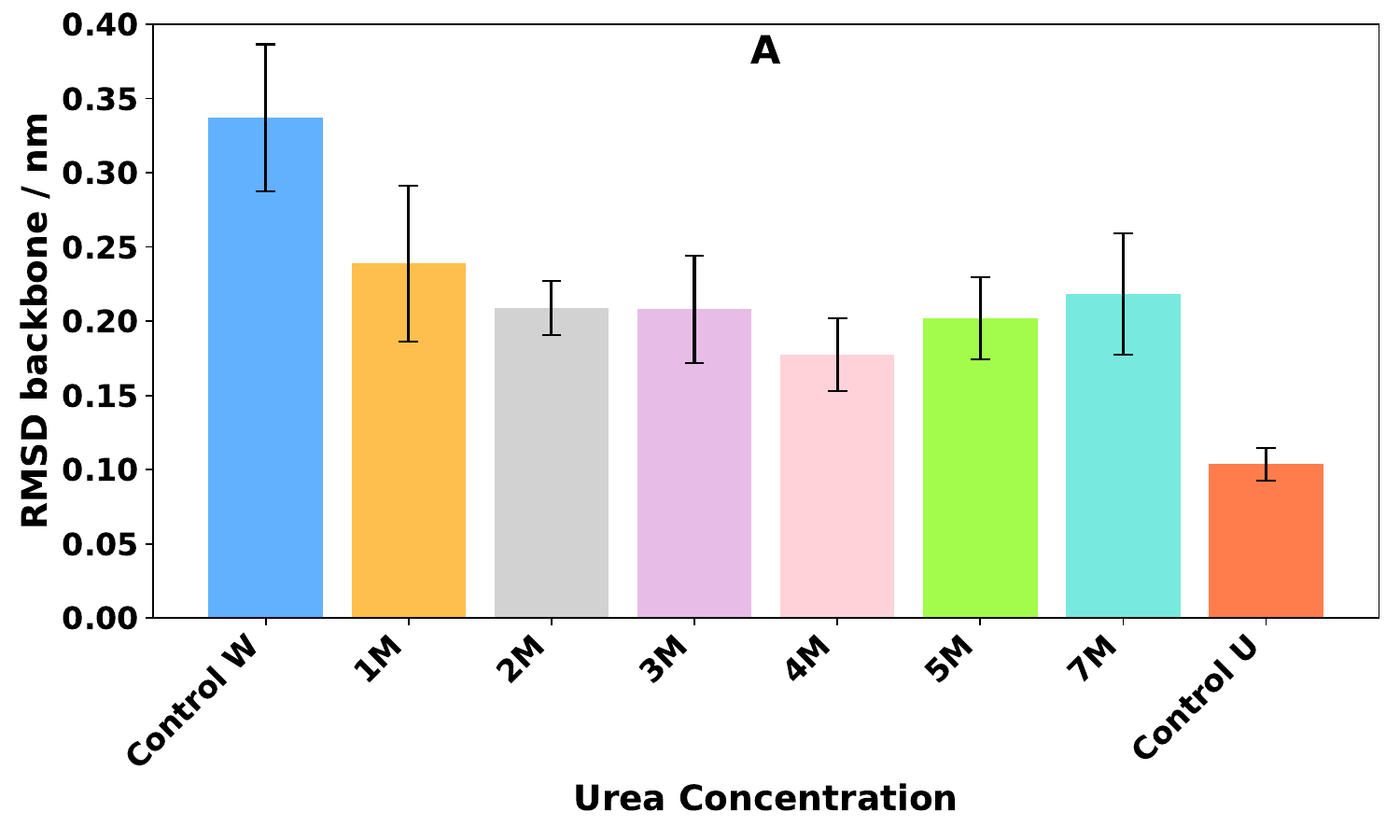}
	\includegraphics[width=0.47\textwidth]{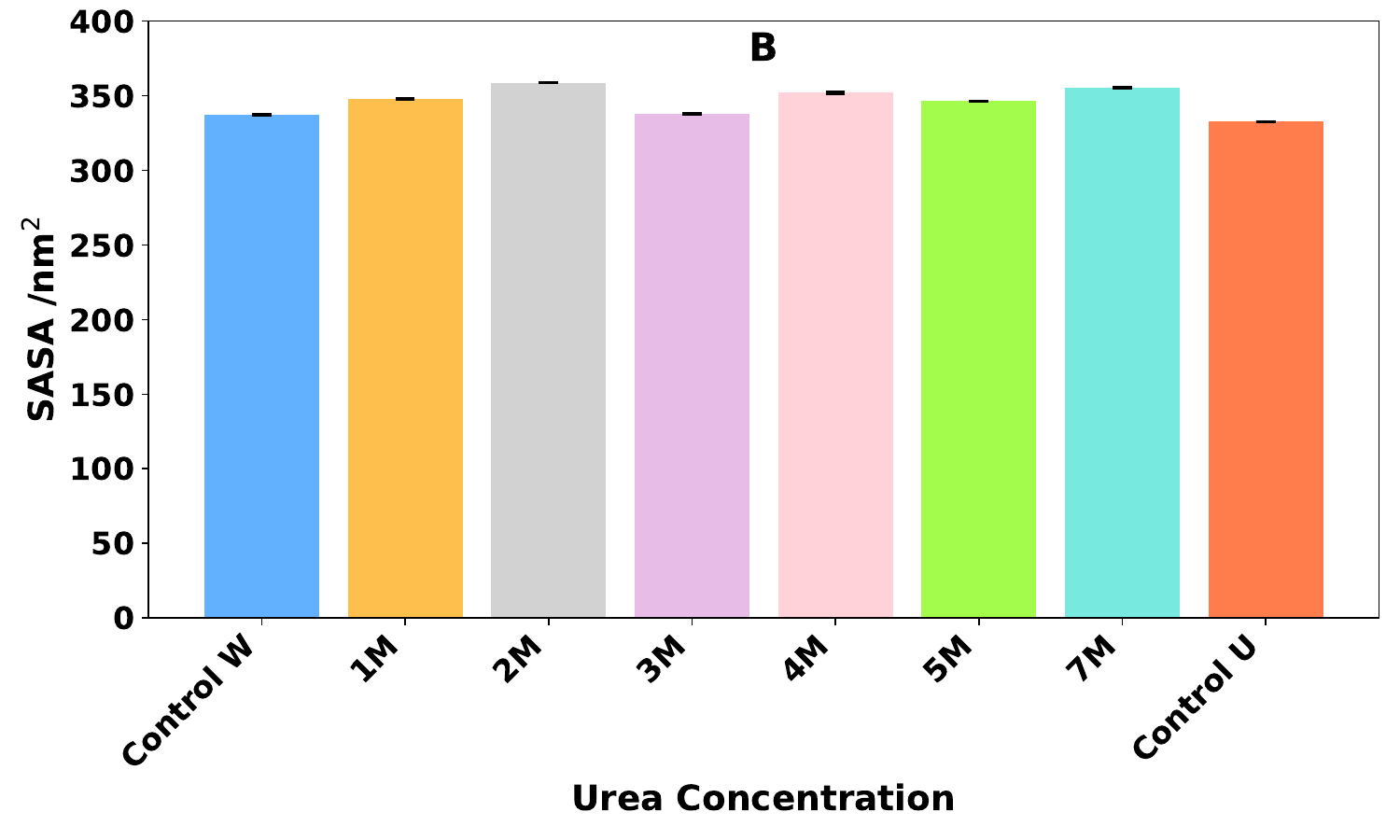}
	\caption{ A- Mean of root mean square deviation vs urea concentration. B- Mean solvent-accessible surface area of the protein vs urea concentration }
	\label{Fig5}
\end{figure}

The analysis of the average RMSD as a function of urea concentration, shown in Figure \ref{Fig5}A, was performed by comparing the structures with the initial configuration of each stage. Relative to the control W system, the lower RMSD values observed in the urea-containing systems indicate reduced deviations from their corresponding initial configurations. However, this behavior should not be interpreted as direct evidence of enhanced thermodynamic stability. Instead, it may reflect a more restricted conformational sampling within the analyzed time window or the fact that the protein starts from an already partially expanded acidic-state conformation.

The SASA analysis shown in Figure~\ref{Fig5}B indicates that the introduction of urea generally increases the solvent-exposed surface area of BSA. In most urea-containing systems, SASA values are higher than those observed in the urea-free control system, except under the complete absence of water (control U). At low urea concentrations, SASA increases monotonically up to 2~M, where the largest average increase is observed. Beyond 3~M, however, this monotonic behavior is lost, and only moderate fluctuations are observed as urea concentration increases. The maximum increase in average SASA is approximately 9\% at 2~M, whereas all other urea-containing systems show smaller deviations relative to the control W system. Moreover, the difference between the pure water and pure urea control systems is only about 2\%, indicating that urea-induced changes in global solvent exposure remain limited in magnitude. This moderate SASA response suggests that the increase in protein--urea hydrogen bonding does not arise from extensive global unfolding, but rather from a remodeling of the protein solvation shell and the exposure of already accessible polar or backbone groups.

\begin{figure}[htbp]
	\centering
	\includegraphics[width=1.0\textwidth]{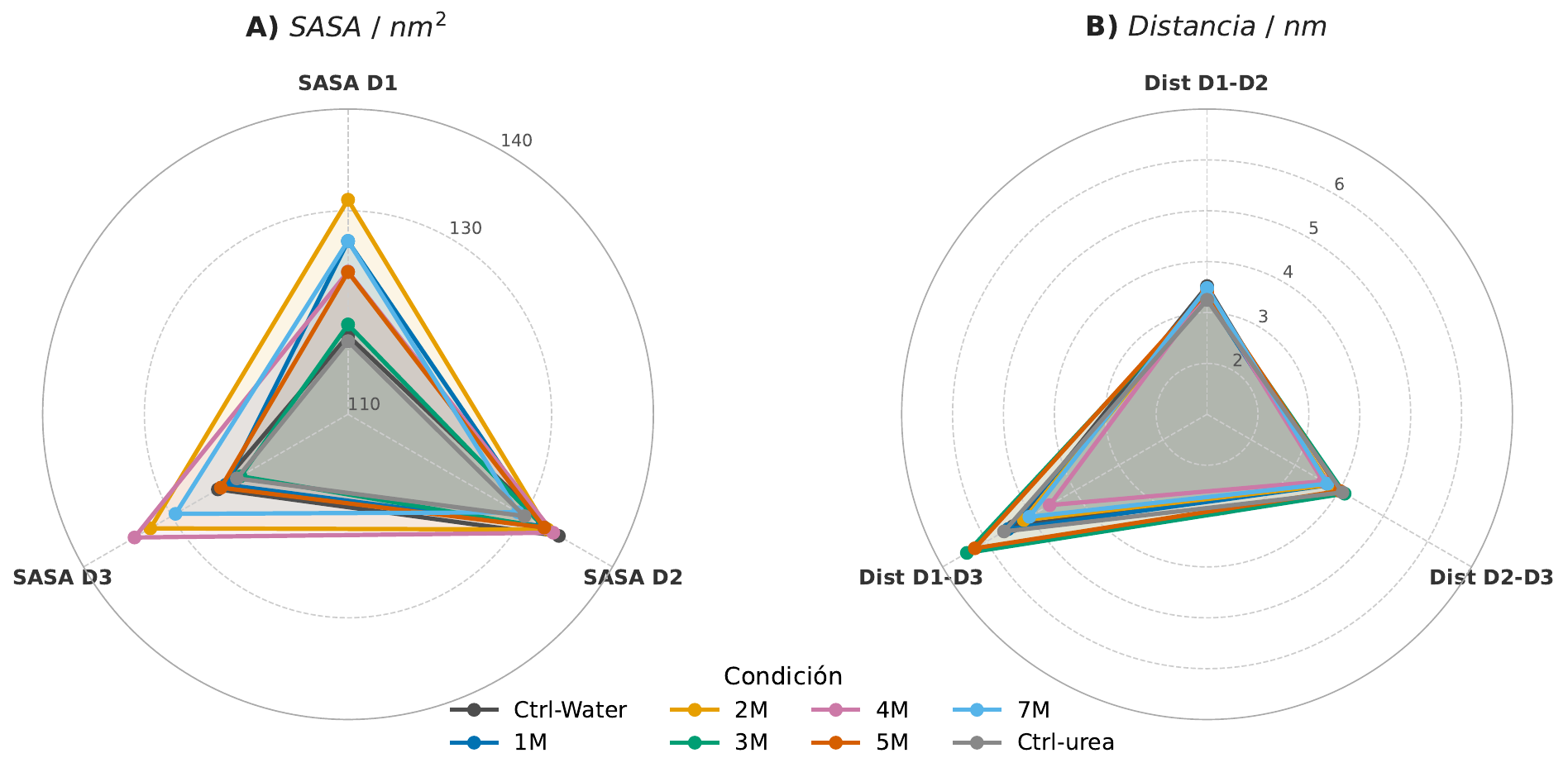}
	\caption{A- Mean solvent-accessible surface area per domain vs urea concentration. B- Mean distance between the center of mass of the domains as a function of
		concentration. The errors in the mean values shown in the figures are small, making them barely visible without enlarging the figure.}
	\label{Fig6}
\end{figure}

Although the changes in global SASA were limited in magnitude, the loss of a clear monotonic trend at higher urea concentrations suggests that solvent exposure may vary differently across individual regions of BSA. To assess this possibility, SASA was analyzed at the domain level, allowing the identification of domains most affected by urea-induced changes in hydration and local solvent accessibility (Figure~\ref{Fig6}A).

The domain-resolved SASA analysis indicates that Domains~I and III exhibit the largest variations in solvent-accessible surface area upon urea addition, whereas Domain~II shows the smallest changes across the concentration range. Domain~I displays a non-uniform response, with increasing SASA values from 0 to 2~M and a general upward trend from 3 to 7~M, followed by a decrease to its lowest value in the absence of water (control U). Despite these variations, Domain~I consistently presents the lowest average SASA among the three domains, suggesting that it remains comparatively less solvent-exposed. However, it also shows the largest relative fluctuations, although these remain below 11\%. Domain~III exhibits slightly smaller SASA variations than Domain~I and does not follow a clear monotonic dependence on urea concentration, with maximum fluctuations of approximately 10\% (Figure \ref{Fig6}A).
The results obtained for Domain~II are consistent with those reported by Scanavachi and co-workers \cite{scanavachi2020aggregation}, who also observed higher and more stable SASA values at nearly all urea concentrations.

To further clarify the role of each domain, we analyzed the interdomain distances, as shown in Figure \ref{Fig6}B.

The interdomain-distance analysis reveals that the D1--D2 separation remains relatively stable across the urea concentration range, indicating that the relative arrangement of Domains~I and II is only weakly affected by the solute. In contrast, the D1--D3 and D2--D3 distances exhibit non-monotonic fluctuations, with higher separations observed at selected urea concentrations, particularly around 3~M and 5~M. These changes suggest that the relative positioning of Domain~III is considerably more variable than that of Domains~I and II. However, the absence of a strictly concentration-dependent increase indicates that urea does not promote a progressive interdomain expansion. Instead, the data support a moderate, domain-dependent rearrangement of BSA, mainly involving the transient or condition-specific displacement of Domain~III. These trends are consistent with the moderate changes observed in the radius of gyration and the domain-resolved SASA analysis, supporting the idea that urea induces local structural adjustments rather than a uniform unfolding or expansion of the entire protein.

\subsection{Preservation of BSA Secondary Structure under Urea Exposure}

To further assess the structural influence of urea, the secondary structure of BSA was evaluated as a function of urea concentration. As shown in Figure~S3 of the Supporting Information, the total number of residues assigned to secondary-structure elements remains nearly unchanged across all conditions, indicating that urea does not induce a substantial loss of secondary structure within the analyzed time window.

A more detailed analysis of individual secondary-structure elements is presented in Figure~S4 of the Supporting Information. There, we observe that the number of residues forming $\alpha$-helices remains largely conserved, with variations of less than 5\% relative to the urea-free control. Given the predominantly helical nature of BSA, this result indicates that the main secondary-structure scaffold of the protein is preserved in the presence of urea. Minor variations are observed in S-bend, hydrogen-bonded turn, and polyproline II conformations, particularly at selected urea concentrations. However, these changes involve a limited number of residues and do not follow a clear concentration-dependent trend.

Overall, the secondary-structure analysis indicates that urea-induced effects on BSA are not associated with extensive loss of helicity or global disruption of secondary structure. Instead, the observed variations are more consistent with small local backbone rearrangements. These findings support the interpretation derived from the hydrogen-bond, SASA, and radius-of-gyration analyses, suggesting that urea primarily reorganizes the protein solvation environment and induces moderate local or interdomain rearrangements rather than complete unfolding.

\subsection{Radial Distribution Analysis of Solvent and Ion Organization}

\begin{figure}[htbp]
	\centering
	\includegraphics[width=0.45\textwidth]{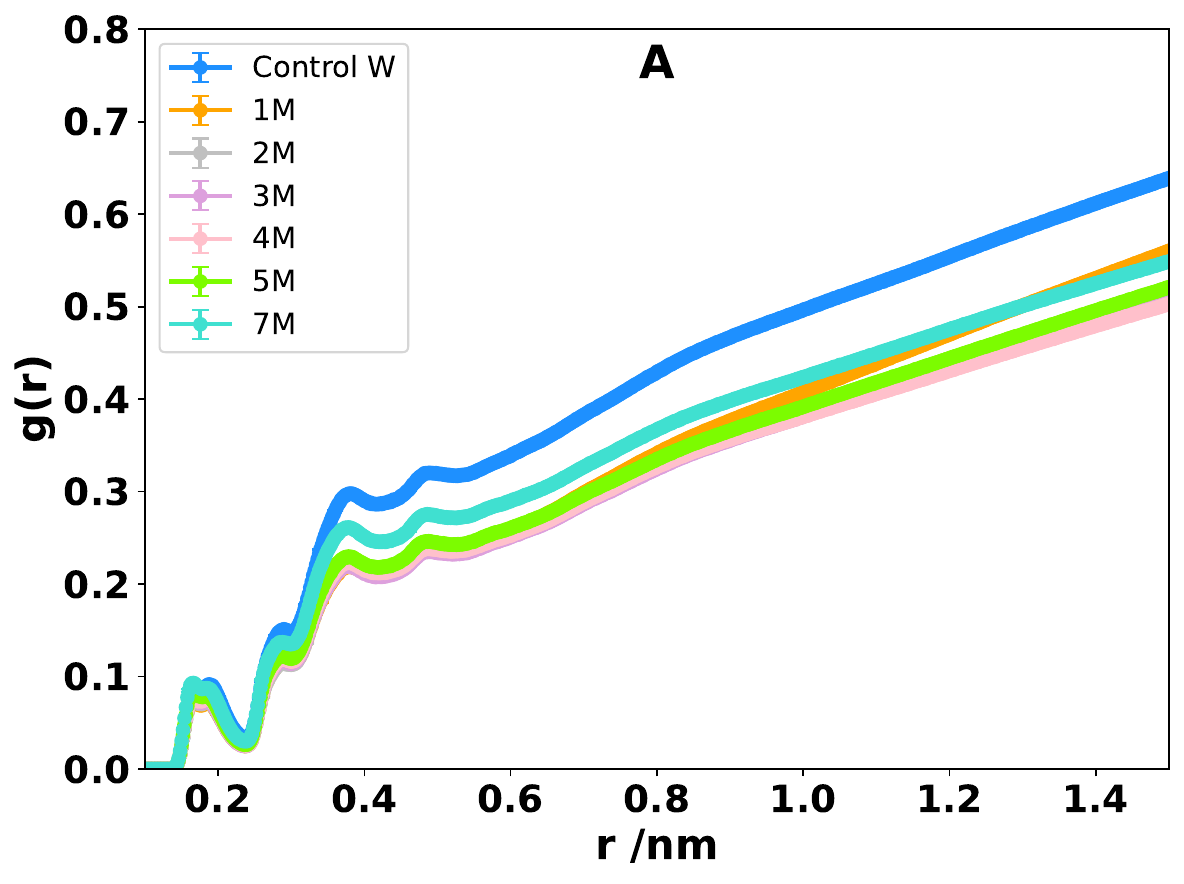}
	\includegraphics[width=0.45\textwidth]{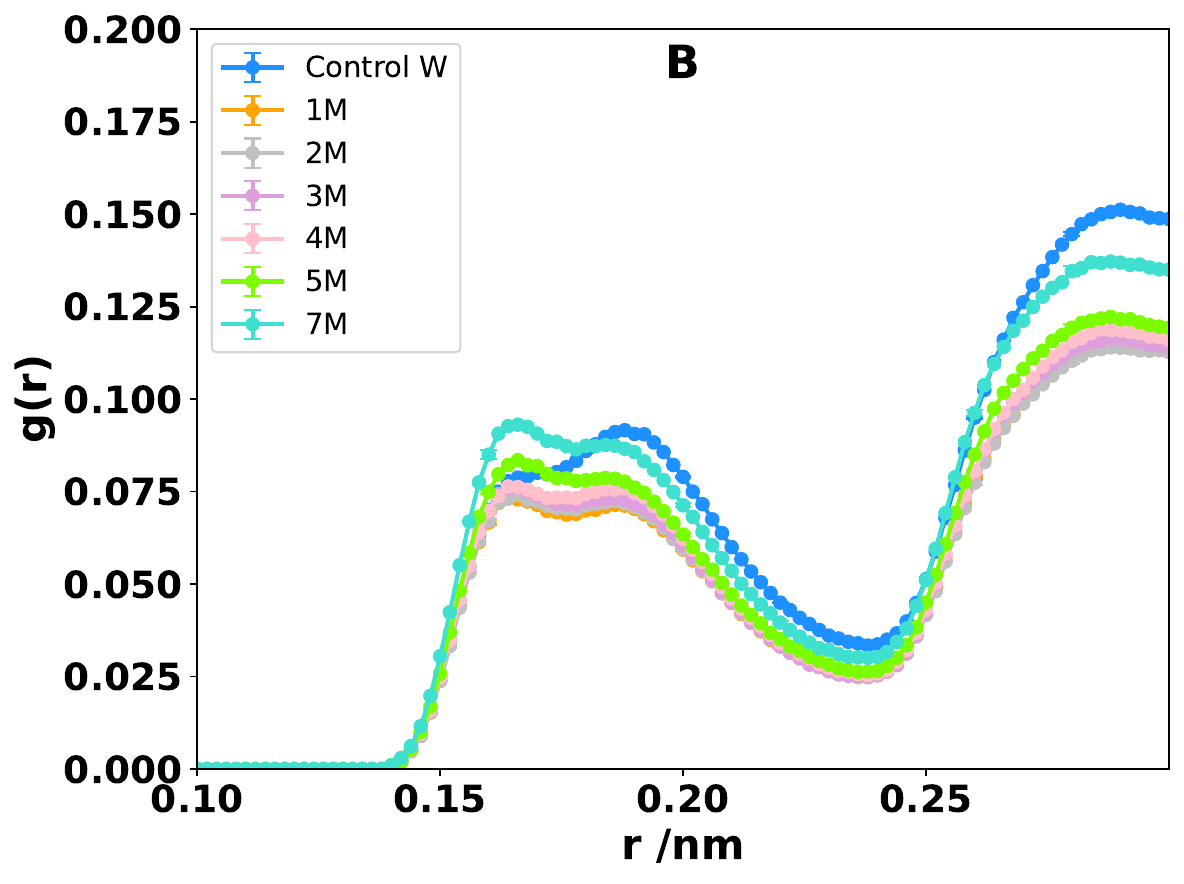}
	\includegraphics[width=0.45\textwidth]{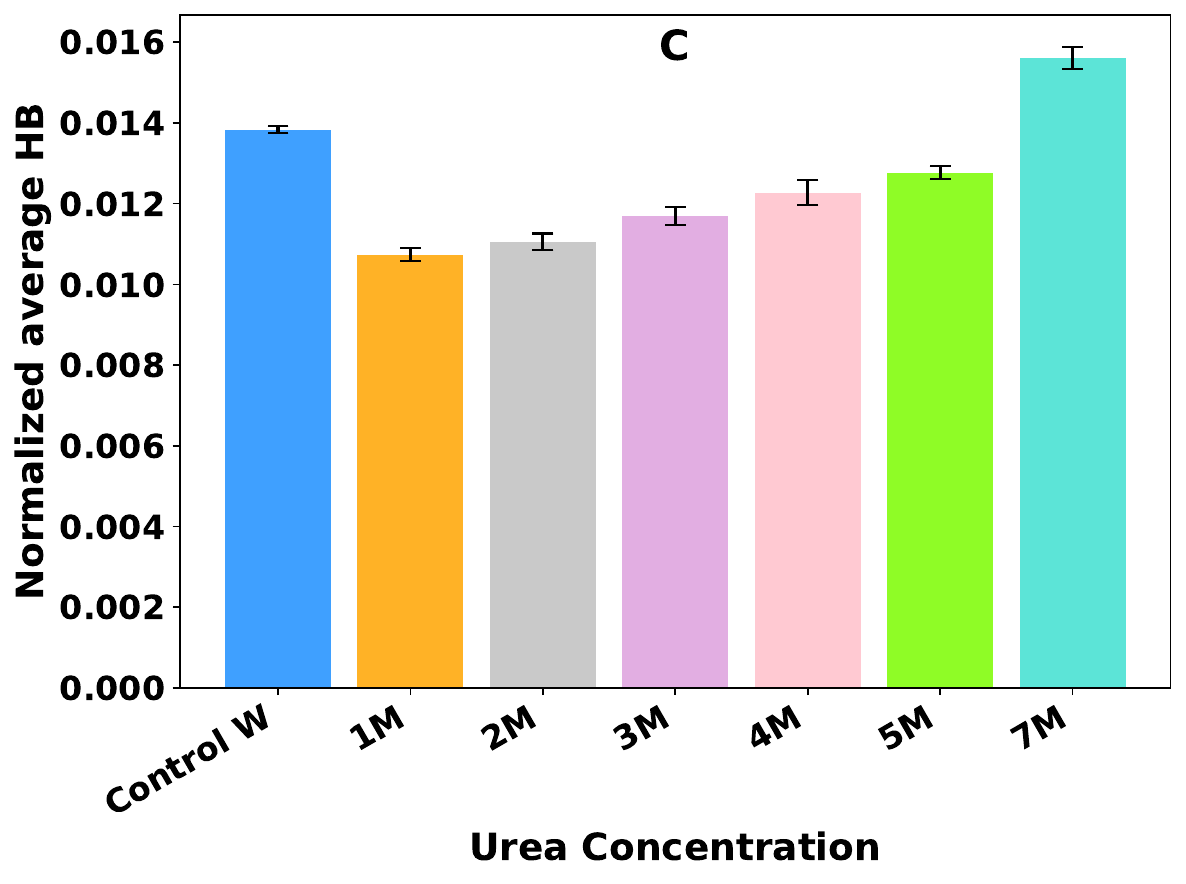}
	\caption{A, B - Radial distribution functions (RDFs) were calculated between the center of mass (COM) of the protein and the COM of water molecules with different ranges. C - The average number of hydrogen bonds (HBs) between protein and water
		molecules is normalized by the number of water molecules.}
	\label{Fig7}
\end{figure}

To further support our initial results and broaden the discussion, we analyzed the radial distribution functions (RDFs) across several groups within the system. In Figure \ref{Fig7}, we show the RDFs calculated between the protein's center of mass (COM) and the water molecules' COM, and separately between the protein's COM and the urea molecules' COM.

In Figure~\ref{Fig7}A, B, the radial distribution function (RDF) between the center of mass (COM) of the protein and the COM of water molecules provides a global description of the radial organization of water around BSA. Because this RDF is referenced to the protein COM, it should not be interpreted as a direct measurement of protein-surface hydration. Instead, it indicates changes in the overall spatial distribution of water molecules around the protein as the urea concentration increases. In the presence of urea, the RDF profiles show lower intensities than the urea-free control over a broad range of distances, indicating a redistribution of water molecules as urea is incorporated into the solvent. Nevertheless, water remains present in the vicinity of the protein across all urea concentrations, suggesting that urea does not completely exclude water from the protein environment. The changes observed in the RDF profiles are not directly proportional to the total number of protein--water hydrogen bonds, but can be better understood by considering the number of hydrogen bonds normalized by the total number of water molecules in each system.

The normalized average number of protein--water hydrogen bonds (Fig.~\ref{Fig7}C) provides a complementary perspective to the RDF profiles shown in Figs.~\ref{Fig7}A and \ref{Fig7}B. From 0 to 1~M urea, the number of protein--water hydrogen bonds per water molecule decreases, reaching a minimum at 1~M. At higher urea concentrations, however, this normalized value increases progressively. This behavior indicates that, although the total number of water molecules decreases as urea concentration increases, the remaining water molecules contribute more efficiently to protein--water hydrogen bonding. Thus, the reduction in total protein--water hydrogen bonds reflects not only the progressive replacement of water by urea, but also changes in the relative participation of the remaining water molecules in the solvation environment of BSA.

\begin{figure}[htbp]
	\centering
	\includegraphics[width=0.45\textwidth]{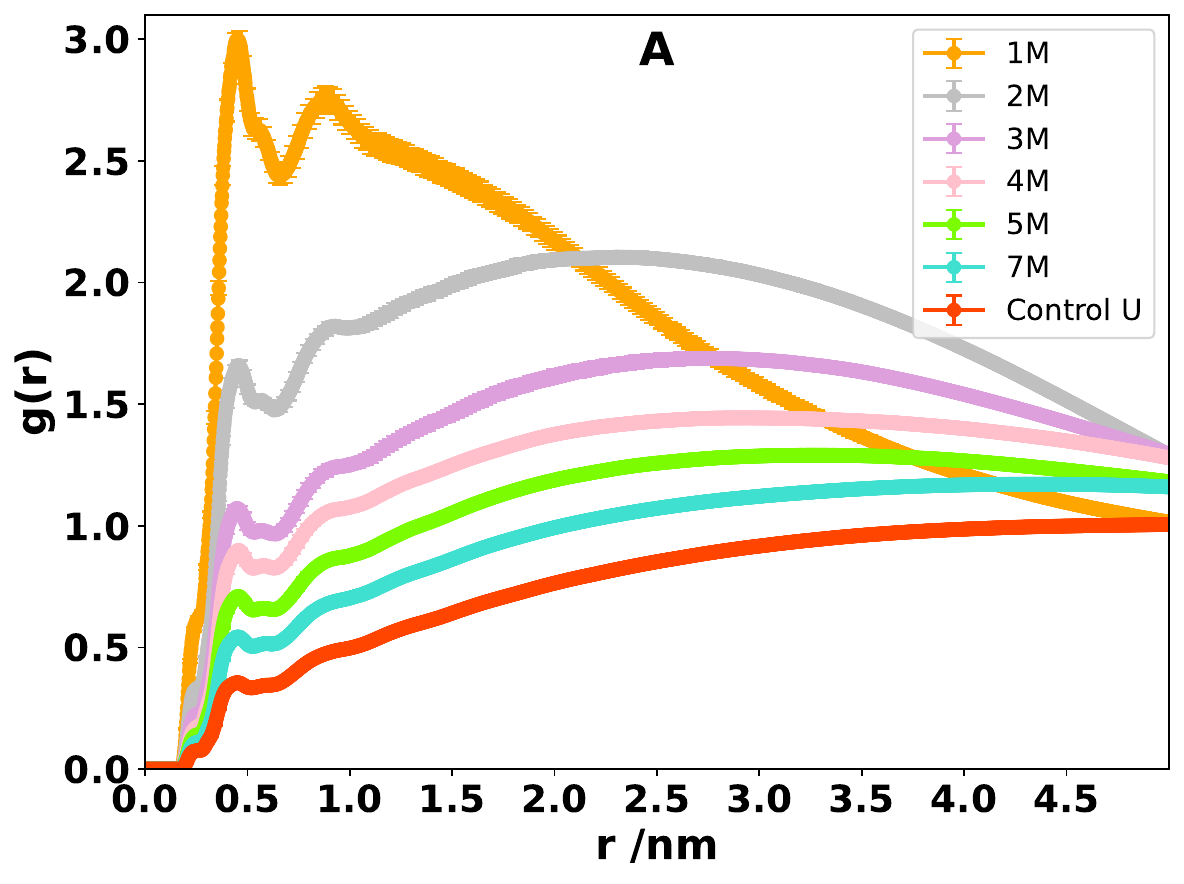}
	\includegraphics[width=0.45\textwidth]{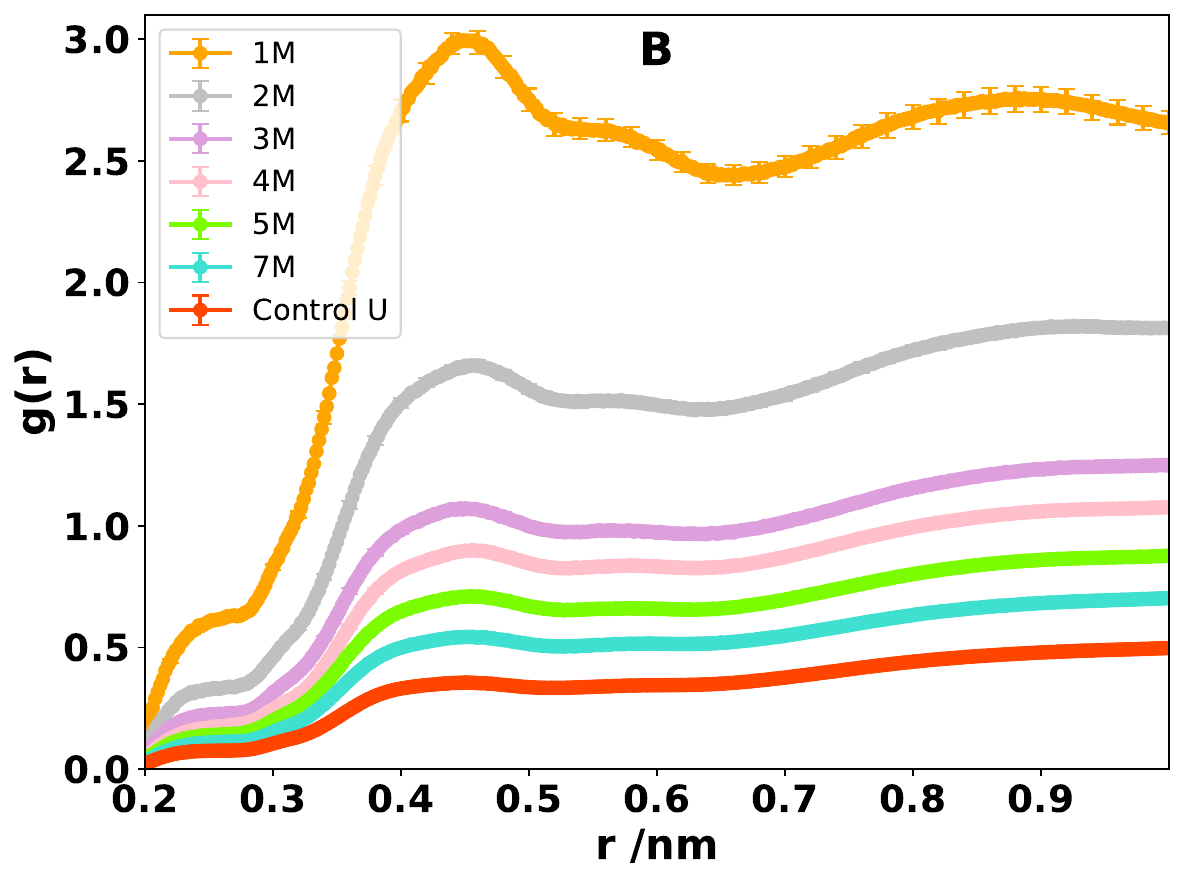}
	\includegraphics[width=0.45\textwidth]{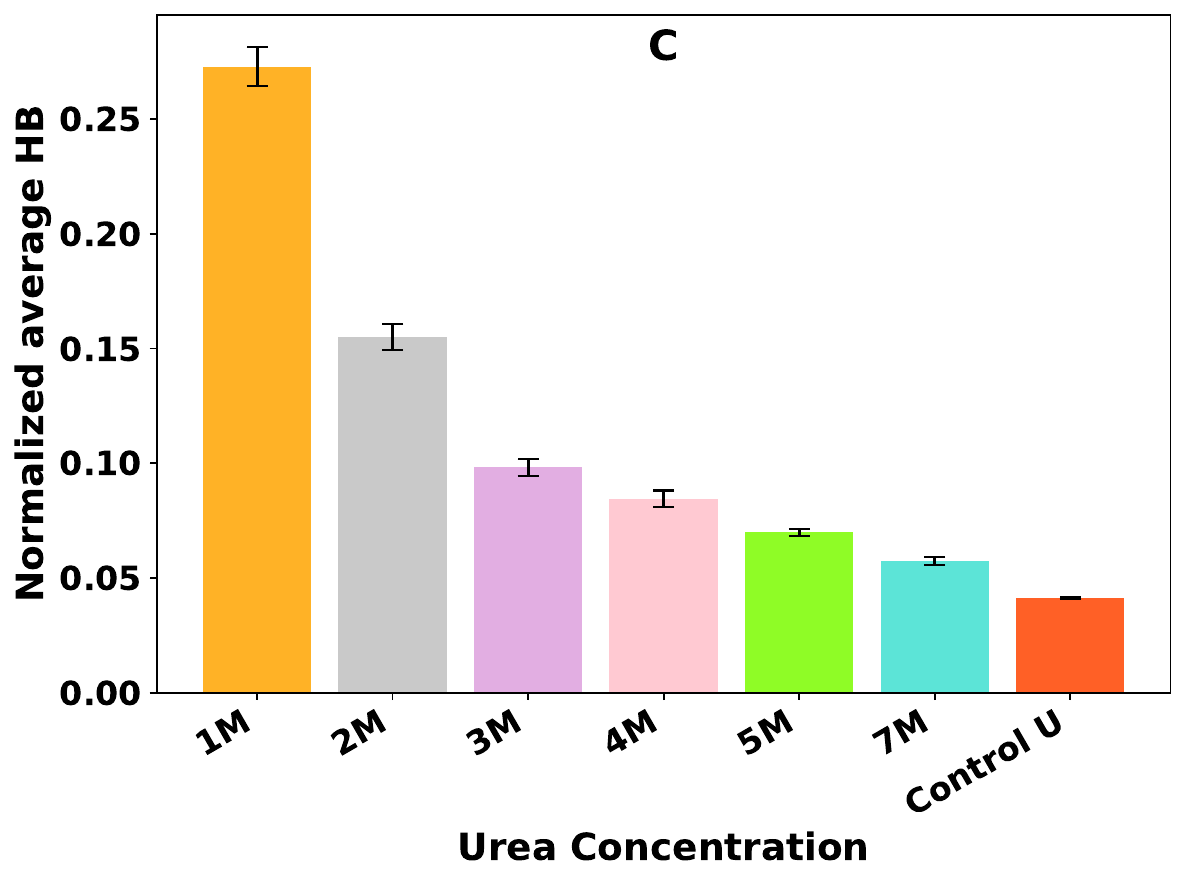}
	\caption{A,B - Radial distribution functions (RDFs) were calculated between the center of mass (COM) of the protein and the COM of urea molecules. C - The average number of hydrogen bonds (HBs) between protein and urea molecules is normalized by the number of urea molecules.}
	\label{Fig8}
\end{figure}

The RDF between urea molecules and the protein, shown in Figure~\ref{Fig8}A, B, displays two distinguishable features at 1~M urea: a first peak around 0.20--0.30~nm and a second feature around 0.40--0.50~nm. These peaks indicate a non-uniform radial distribution of urea around the protein. As urea concentration increases, the peak intensities decrease; however, because RDFs are normalized by the bulk density of urea, this trend should not be interpreted as a decrease in the absolute number of protein-associated urea molecules. Instead, it suggests reduced relative enrichment of urea near the protein at higher concentrations, likely reflecting redistribution of urea within the solvent environment.

The analysis of protein--urea hydrogen bonds normalized by the number of urea molecules (Fig.~\ref{Fig8}C) provides a complementary perspective to the total hydrogen-bond analysis. Although the total number of protein--urea hydrogen bonds increases with urea concentration, the normalized average number of hydrogen bonds per urea molecule decreases across the concentration range. This behavior indicates that, as the bulk concentration of urea increases, the relative contribution of each urea molecule to direct protein binding becomes lower. Therefore, the increase in total protein--urea hydrogen bonding is mainly driven by the larger number of urea molecules available in the system, rather than by an increased binding efficiency of each urea molecule. This result also suggests that water continues to participate in the protein solvation environment, while excess urea may increasingly redistribute within the bulk solvent or engage in urea--urea interactions.

\begin{figure}[htbp]
	\centering
	\includegraphics[width=0.45\textwidth]{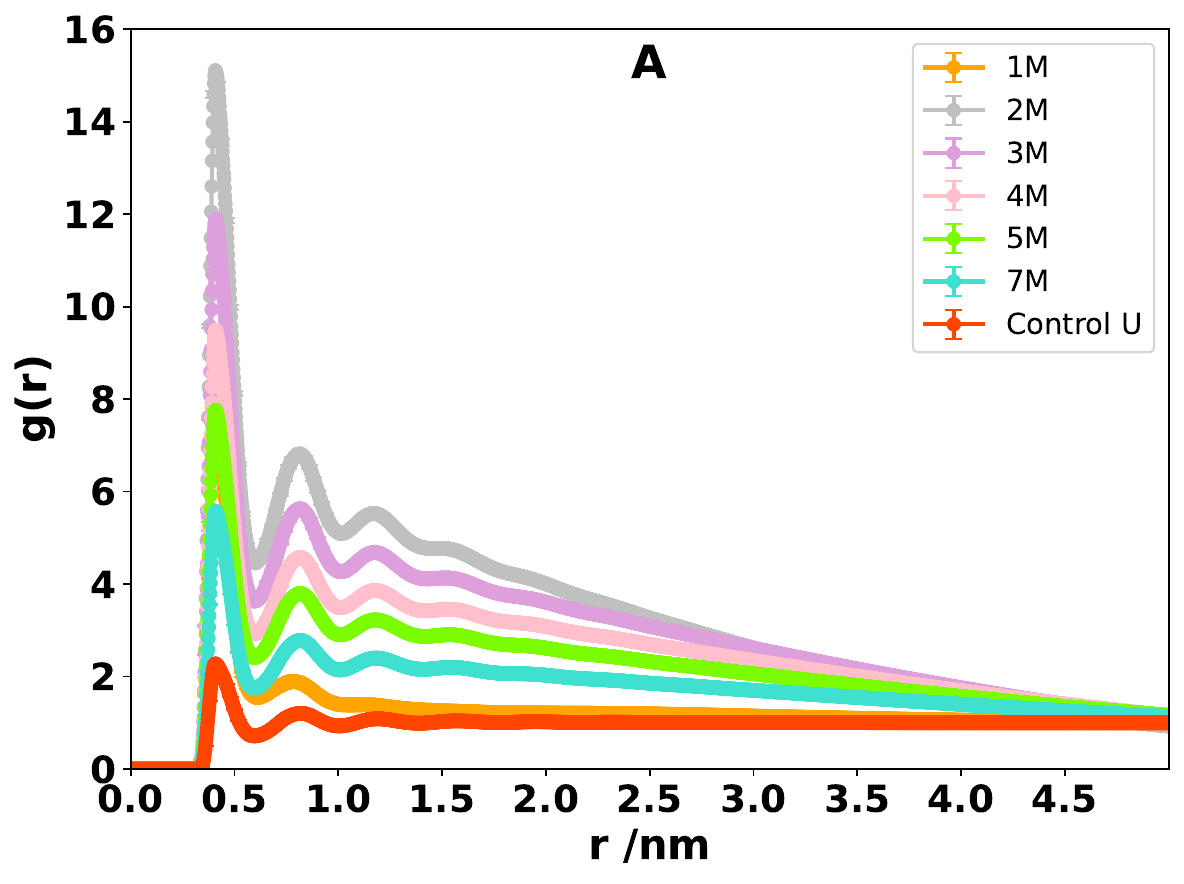}
	\includegraphics[width=0.45\textwidth]{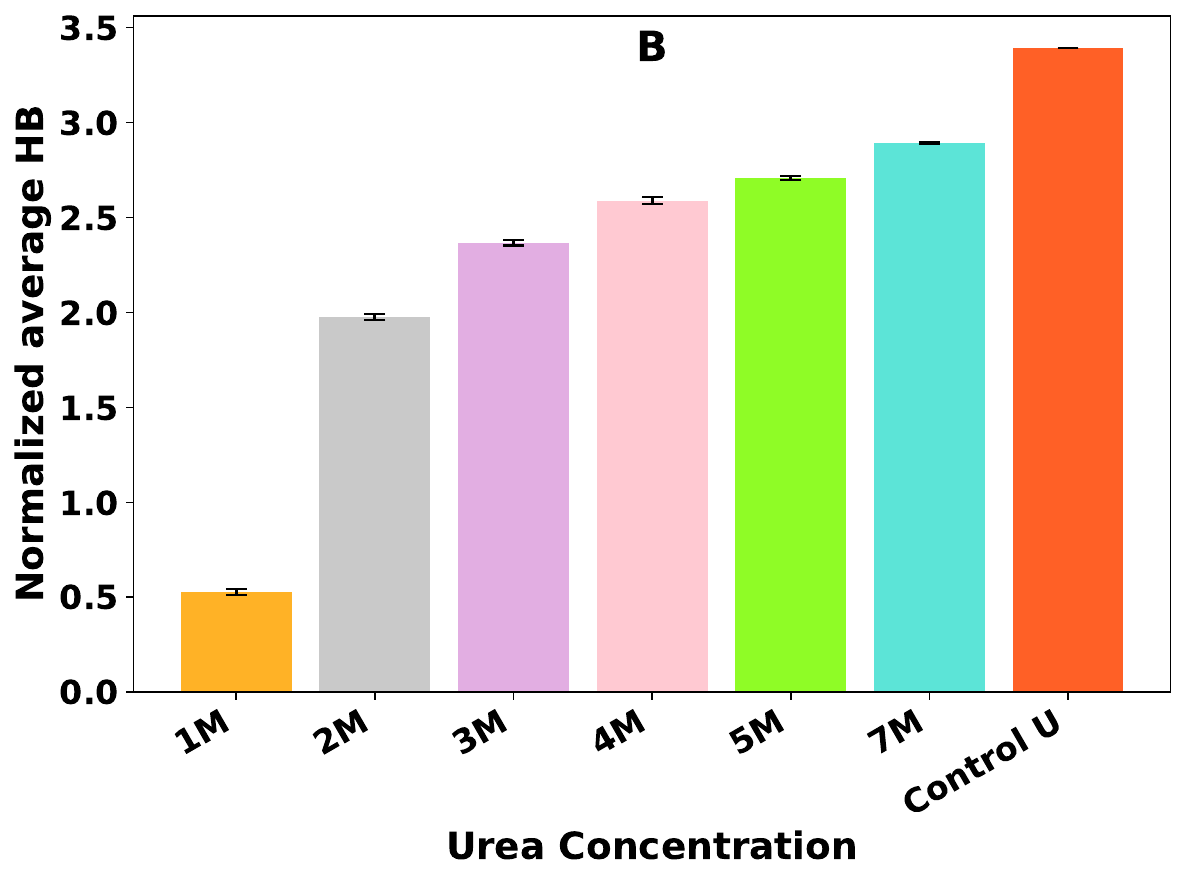}
	\caption{A - RDF between the carbon atoms of urea molecules at different urea concentrations. B - The average number of hydrogen bonds (HBs) between urea molecules normalized by the number of urea molecules.}
	\label{Fig9}
\end{figure}

The radial distribution function between the carbon atoms of urea molecules was analyzed to evaluate urea--urea organization in solution (Fig.~\ref{Fig9}A). The increase in the intensity of the first and second coordination peaks at low to intermediate concentrations, particularly between 2 and 4~M, suggests enhanced local self-association among urea molecules. Therefore, this RDF should be interpreted as evidence of urea--urea structuring in the solvent, rather than structuring around the protein. This behavior may reflect specific geometric arrangements adopted by urea molecules during transient aggregation, in agreement with previously proposed models of urea self-association \cite{atahar2019aggregation}.

The normalized urea--urea hydrogen-bond analysis (Fig.~\ref{Fig9}B) is consistent with the RDF profiles, indicating that urea self-association increases with concentration. This suggests that, at higher urea concentrations, urea--urea interactions contribute increasingly to solvent organization.

The RDF between chloride ions and water molecules was analyzed to evaluate how urea modifies the local ionic solvation environment (see Figure~S5 in the Supporting Information). A main peak is observed around 0.28--0.34~nm, consistent with the first solvation shell of chloride ions by water molecules. In the presence of urea, changes in the intensity of this peak indicate that the chloride hydration environment is altered as the solvent composition changes. However, these variations should not be interpreted as a direct loss of water--chloride interactions without further coordination-number analysis. Rather, they suggest a redistribution of water molecules around chloride ions associated with the progressive replacement of water by urea in the bulk solvent.
The RDF between the carbon atom of urea molecules and chloride ions was analyzed to evaluate whether changes in chloride hydration are accompanied by a redistribution of chloride toward urea (see Figure~S5 in the Supporting Information). The resulting profiles show that chloride--urea organization is modified as the solvent composition changes; however, no simple concentration-dependent trend is observed across all systems. The Control U system exhibits distinct behavior, likely reflecting the absence of water and the resulting redistribution of chloride ions in a purely urea-based environment. The RDFs between the protein center of mass and chloride ions (see Figure~S6 in the Supporting Information) further indicate that chloride distribution around BSA is strongly affected only under the extreme Control U condition. In the urea--water mixtures, the profiles are broadly similar to the Control W system and do not show a clear monotonic dependence on urea concentration. Therefore, chloride-mediated interactions appear to play a secondary role in the structural response of BSA to urea, compared with the dominant solvent-remodeling effects associated with protein--water, protein--urea, and urea--urea interactions.

Although ion--solvent interactions have been discussed as relevant contributors to urea solution structure by Espinosa and co-workers \cite{espinosa2018mechanisms}, the present RDF analysis does not indicate a central role for urea--chloride or water--chloride interactions in the structural response of BSA to urea. Instead, the dominant effects appear to be associated with solvent-shell remodeling involving protein--water, protein--urea, and urea--urea interactions.

\section{Conclusion}

This study employed molecular dynamics (MD) simulations to investigate the effects of urea on the structural properties of bovine serum albumin (BSA) at pH 3.7. Under these conditions, BSA adopts a partially unfolded conformation characterized by increased solvent exposure of its domains, commonly referred to as the F isoform. The simulated systems were constructed over a broad range of urea concentrations, from 0 M (water control, Control W) to a fully urea-solvated system containing urea as the sole solvent component (urea control, Control U).

Within this concentration range, the calculated radius of gyration (RG), root-mean-square deviation (RMSD), solvent-accessible surface area (SASA), and interdomain distances indicate that urea-induced structural changes in BSA remain moderate at the global level but become more pronounced in the local solvation environment and domain organization. This behavior is consistent with the hydrogen-bond analysis, which shows that urea progressively replaces water molecules within the protein solvation shell without promoting large-scale protein expansion. The modest increase in SASA and the limited variations in RG suggest that the increase in protein--urea hydrogen bonds primarily reflects solvent-shell reorganization and the exposure of accessible polar and backbone groups rather than complete unfolding. In addition, the nonmonotonic variations observed in the D1--D3 and D2--D3 distances suggest that Domain~III undergoes moderate, condition-dependent interdomain rearrangements that may facilitate local changes in hydration and urea binding.

The secondary-structure analysis reveals that the effects induced by urea are not associated with extensive helix loss or large-scale disruption of the protein secondary-structure framework. Instead, the observed structural variations are consistent with subtle local backbone rearrangements. These findings are in agreement with the hydrogen-bond, SASA, and radius-of-gyration analyses, collectively suggesting that urea primarily modulates the protein solvation environment while promoting moderate local and interdomain conformational rearrangements rather than complete unfolding.

Radial distribution function (RDF) analysis reveals persistent local hydration near solvent-accessible regions of the protein core across all urea concentrations, with evidence of partial rehydration at higher concentrations. Consistently, the normalized (per-water-molecule) protein--water hydrogen-bond analysis indicates that the remaining water molecules engage in hydrogen bonding with the protein more efficiently at elevated urea concentrations; this reflects a relative gain in per-molecule hydrogen-bonding efficiency rather than a net increase in the absolute number of hydration water molecules. Increasing urea concentration is accompanied by a redistribution of water molecules around chloride ions, although this trend should not be interpreted as a direct, monotonic promotion of water--chloride interactions; chloride interactions with urea or the protein remain comparatively limited except under water-depleted conditions. In parallel, urea--urea interactions become increasingly favorable, with urea self-association dominating at high concentrations.

A key feature emerging from the analysis is the progressive dehydration of the protein surface, {\color{red}in absolute terms,} with increasing urea concentration, driven by the replacement of protein--water hydrogen bonds with protein--urea interactions. Although this exchange mechanism maintains or slightly increases the total number of protein-associated hydrogen bonds, comparison between the two control systems indicates that urea does not fully restore the original protein--water hydrogen-bond network. This incomplete compensation may contribute to the observed local structural destabilization. Domain-level analysis further supports the existence of a dynamic compensation mechanism in which the total number of hydrogen bonds involving the protein, water, and urea remains nearly constant.

Urea predominantly disrupts hydrogen bonding between the protein backbone and water, producing an approximately 40\% reduction in backbone--water hydrogen bonds across all concentrations. Simultaneously, the increase in backbone--urea hydrogen bonds indicates preferential urea binding and supports a competitive solvation mechanism in which urea progressively replaces water at key hydrogen-bonding sites, potentially facilitating conformational destabilization. Analysis of side-chain interactions further corroborates this compensation mechanism, showing that urea partially substitutes for water at accessible polar and charged residues while preserving the local interaction network.

\begin{acknowledgement}

	The authors thank and fondly remember Raúl Grigera, who
	taught and shaped us in this beautiful profession. We also thank
	the anonymous reviewers for their thoughtful feedback on earlier
	drafts of this article. This work was supported by grants from the
	Consejo Nacional de Investigaciones Científicas y Técnicas
	(No. KE3-11220210100918CO) and the National University Arturo
	Jauretche (No. 80020230100029UJ)
\end{acknowledgement}
\begin{suppinfo}
	The following files are available free of charge.
	\begin{itemize}
		\item Figures~S1 and~S2: backbone-- and side-chain--resolved hydrogen bonds with water and urea.
		\item Figure~S3: total number of residues assigned to secondary-structure elements vs.\ urea concentration.
		\item Figure~S4: residue counts per individual secondary-structure element vs.\ urea concentration.
		\item Figure~S5: RDFs of chloride--water.
        \item Figure~S6: RDFs of chloride--urea.
		\item Figure~S7: RDFs between the protein center of mass and chloride ions.
	\end{itemize}
\end{suppinfo}

\section*{Data Availability}
The simulation input files, topologies, and all available material to reproduce all trajectories that support the findings of this study are available in the following repository https://github.com/MEL-IFLYSIB/BSA-37.git.

\bibliography{bsa37ref}
\end{document}